# The Dark Web Phenomenon: A Review and Research Agenda

*Full Paper*

**Abhineet Gupta**
School of Computing and Information Systems
University of Melbourne
Melbourne, Australia
Email: abhineetg@student.unimelb.edu.au

**Sean B Maynard**
School of Computing and Information Systems
University of Melbourne
Melbourne, Australia
Email: seanbm@unimelb.edu.au

**Atif Ahmad**
School of Computing and Information Systems
University of Melbourne
Melbourne, Australia
Email: atif@unimelb.edu.au

## Abstract (Abs heading)

The internet can be broadly divided into three parts: *surface*, *deep* and *dark*. The *dark web* has become notorious in the media for being a hidden part of the web where all manner of illegal activities take place. This review investigates how the dark web is being utilised with an emphasis on cybercrime, and how law enforcement plays the role of its adversary. The review describes these hidden spaces, sheds light on their history, the activities that they harbour – including cybercrime, the nature of attention they receive, and methodologies employed by law enforcement in an attempt to defeat their purpose. More importantly, it is argued that these spaces should be considered a phenomenon and not an isolated occurrence to be taken as merely a natural consequence of technology. This paper contributes to the area of dark web research by serving as a reference document and by proposing a research agenda.

# 1  Introduction

The Internet provides a platform for an information system, known as the World Wide Web, where information can be exchanged by a global community of citizens. Within this community exists the Dark Web or Darknet, an environment that affords its users anonymity, making attribution for activities challenging by encrypting and routing users' traffic via multiple nodes ([The Tor Project 2018](#)). The most popular version of the dark web, The Onion Routing (Tor) network and protocol, has become a haven for criminal activity, including sharing illegally-acquired information, trading illicit contraband, and recruiting others, all with disregard for borders and legality ([Dalins et al. 2018](#); [Vogt 2017](#)). The dark web became popular with the launch of *Silk Road* – a drug marketplace – in 2011, and also by its demise in 2013 ([Van Buskirk et al. 2014](#)). Ever since, the *dark web* has been notorious for facilitating illegal activities and is increasingly being targeted and monitored by authorities.

The motivation behind this paper is to <u>gauge the current state and growth of the dark web</u> in relation to the roles it plays, <u>investigate how the dark web enables cybercrime</u>, and <u>examine law enforcement's efforts</u>. More specifically, the following research questions are under focus:

RQ1. What are the roles that the dark web plays in society?

RQ2. What significance does the dark web have in cybercriminal activities and operations?

RQ3. How successful is law enforcement in its attempts to curb illegal activity on the dark web?

In this paper, we first introduce the area of the dark web and then discuss our literature search methodology. We then discuss the literature in terms of the dark web as a phenomenon, the major roles the dark web plays and discuss legal and societal concerns. Lastly, we present a research agenda in the form of a set of research questions and conclude the paper.

# 2  Background

The internet has enabled the formation of a global digital society that transcends boundaries, be they nationalities, legal jurisdictions, race, or religion. This society, despite its amorphous nature, still constitutes individuals bound by the laws of their country ([Papacharissi 2002](#)). This is possible because online identities, mainly IP addresses, are linked to individuals or websites who possess them. With this level of accountability, attribution for most activities conducted online is easy to perform, especially for law enforcement, as they can monitor traffic or request logs from those who possess them. Therefore, the idea of digital anonymity (lack of attribution between a digital identity and a physical one) is important and forms the basis of the following discussion.

Using the attributes of *public* vs *private* and *accountable* vs *anonymous the internet can be* divided into three broad categories ([Lautenschlager 2016](#)). 1) the *surface web*: is *public* because access is not restricted by authentication or payment, is indexed by search engines, and is *accountable* as the stakeholders are identifiable thus subject to law enforcement. 2) the *deep web:* is parts of the internet not publicly accessible (i.e. *private*), is not indexed by search engines, and is *accountable.* Access is restricted due to authentication requirements or because it forms part of an internal network. *Accountability* in some ways is even stronger than on the *surface web* given authentication requirements. 3) the *dark web:* also known as *darknets* or *hidden services* ([Broadhurst et al. 2017](#))*,* is a subset of the internet and is not indexed by search engines because it requires the use of special software for access. It consists of both *public* and *private* elements (i.e. accessible publicly or by only those with credentials) provided the correct software is in use. The key difference between the dark web and surface or deep web lies in the lack of *accountability* ([Lautenschlager 2016](#)). Users are unidentifiable to the network, or anyone monitoring, and their actions are thus effectively anonymised. Furthermore, the dark web allows for hosting of web services (hidden services) which remain anonymous with regards to their true IP address, and thus location, even to the users who use those web services. By conferring anonymity, private engagements between people have been institutionalised by the dark web.

The most prominent manifestation of the dark web (the Tor network) routes internet traffic via multiple nodes, each only aware of the sender and destination in their immediate vicinity and thus unaware of the original sender and destination of that traffic. Like the Internet, the dark web exists on a system which is decentralised in nature with no central servers or point of control, so it is difficult to shut down ([Tanenbaum and Van Steen 2007](#)). Organisations, such as human-rights activist groups or universities promoting open access, are known to actively donate bandwidth to this cause and run relays ([Owen and Savage 2016](#)). The Tor protocol uses an encrypted connection to each destination across multiple nodes (or relays) which enables the Tor browser to provide anonymity even when browsing to websites on the surface web. This system, when combined with public-key cryptography and the layering of traffic in such a way that for each connection a node is only aware of its neighbouring nodes, means that any one participant in the network never has complete knowledge of the identity of any end-to-end communication channel. For many users, the prime reason for using an overlay network like Tor is not just for anonymous access to regular websites, but to access a range of websites that are not otherwise accessible on the surface web, except for only via the Tor network (Dalins et al. 2018).

## 3   Research Method

A major initial challenge of performing a literature review includes identifying and structuring the most relevant literature surrounding the topic. The primary approach taken for this included performing a keyword search with filters in major databases, going backwards and forward, i.e. reviewing citations for key articles, and reviewing material which cites those key articles, respectively (Webster and Watson 2002). The two primary sources used for this review were *EBSCOHost* and *Google Scholar*. The identification process, described below, was adapted from (Mathiassen et al. 2007). The broad keywords used for the searches initially included "dark web" or "darknet" as these represented the main idea under investigation. Once Tor was determined to be the most popular version of the dark web, "tor" and "tor hidden services" were added to the keyword list. The search was also restricted to contain articles only published in the last five years, i.e. 2013-current. Due to the results being so broad in nature, a random selection of papers was initially reviewed to identify the key themes of research. From these, the themes of interest were then searched for specifically along with the main keyword ("dark web") by adding keywords like "markets", "cybercrime", future", "tor hidden services", "incident response", "information warfare" and "threat intelligence". These additional keywords were used to allow a more detailed discussion on each aspect of the topic and 151 journal articles, books and theses were identified. The list was used to populate a concept matrix based on a brief reading of each article. Papers were then manually filtered for stronger relevance by selecting only those which focused on the dark web and not merely considered it as a peripheral. This resulted in 41 articles in the final list. Many of these articles are centric to the USA and thus may bias the feel of the discussion in this paper towards those sources (for example the discourse around civil liberties).

Overall, there is limited availability of research focusing on the dark web. This may be due to the restrictive nature of the topic which makes it difficult to collect data. By nature, Tor Hidden Services are not indexed on a search engine which makes it necessary to manually compile a list of '.*onion*' addresses before they can be crawled. Additionally, users of the dark web are by nature anonymous which makes it difficult to collect data about them. While the Tor Metrics project does present numerical data around Tor's usage, details of the traffic flows remain unknown. These two limitations: lack of collectible metadata and unindexed web content, in combination with specialised software required to access the network, make the dark web not a well-researched, or published, topic in academic literature.

## 4   Discussion of Literature

The dark web is a feature of a specific overlay distributed system existing atop the global internet that provides the functionality of remaining untraceable and promises to be a haven from prying eyes, mainly law enforcement. Technological improvements have allowed for novel ways to enable the dark web to exist and sustain itself longer than before.

### 4.1   Dark Web as a Phenomenon

The dark web's original offer of anonymity and lack of attribution to a real-world entity makes it a lucrative place for activity that would otherwise be hindered by government control (Weimann 2016a). There are many ways in which this aspect of the dark web manifests itself. These include having marketplaces for illegal contraband; partaking in social interactions involving training, recruitment and propaganda on controversial topics; and conducting and facilitating financial transactions without a paper trail (Chertoff and Simon 2015). All of these are features typical of a black market economy which in itself is a consequence of the market phenomenon of parties engaging in exchange (Ablon et al. 2014). The existence of a digital space which encompasses the above functions, people and technologies could suggest that the dark web is a *phenomenon* arising from a need for secrecy and anonymity between members of a community, similar to that of a physical-world black market existing from a need for unregulated exchange of goods and services between individuals.

Also, like a black market, the dark web's hidden services, or at least those with illicit content, are the main reason behind the dark web's adversarial relationship with law enforcement. Law enforcement attempts to shut down certain hidden services or to deanonymize its users while at the same time participants in this phenomenon find ways to mitigate those attempts. The common goal of anonymity and trading has brought together people in a sort of global private space that is indifferent, and immune to an extent, to local jurisdictions in which their participants reside. The bestowment of anonymity on willing global citizens has translated into a market-based society of its own which tries its best to resist outside efforts of disruption.

## 4.2  Major Roles of the Dark Web

The promise of anonymity on the dark web opens itself up to legitimate uses (e.g. civilians looking for protection from irresponsible corporations, militaries hosting hidden command and control services, journalists conducting operations in countries without access to free media and speech; law enforcement performing sting operations) as well as illegal activities (e.g. command-and-control endpoints for botnets, markets allowing zero-day exploits to be traded, hackers for hire, private communication, coordination for attacks, variable-sized botnets for hire to launch Distributed Denial of Service attacks and pawning of stolen information (Winkler and Gomes 2016)). Legitimate uses may necessitate use of the dark web in some countries while not necessary in others (Chertoff and Simon 2015) due to the variation in the legal landscape across nations. Furthermore, individuals who wish to remain anonymous in the physical world can have a persona which represents them across multiple services online. They can have a stable and consistent presence online while at the same time be physically mobile and always on the move, something that is a necessary extra step in avoiding legal prosecution in the physical world, especially in the highly collaborative nature of developed countries. Figure 1 shows the topics about which Tor hidden services (at least those with a website front) exist (size of a node is proportional to its topic presence, and weight of an edge expresses relatedness of the nodes) (Spitters et al. 2014).

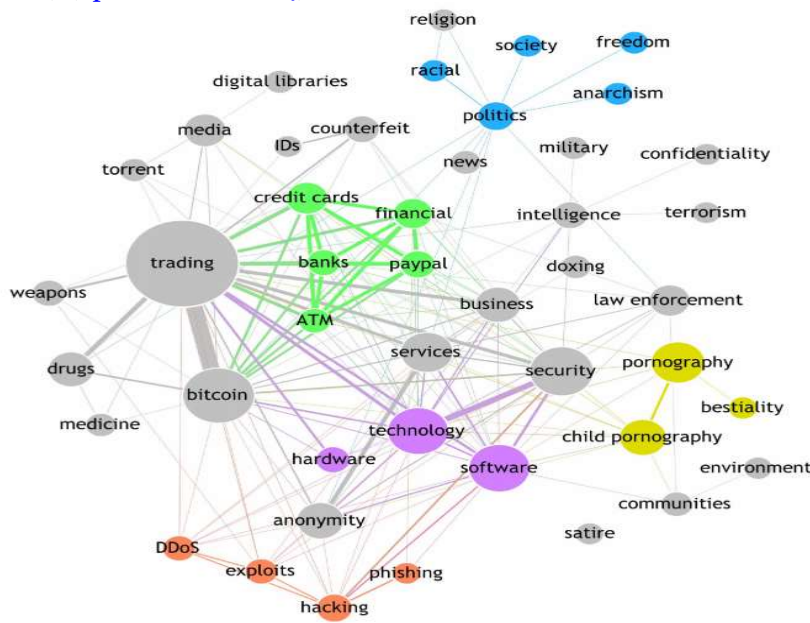

*Figure 1: A topic taxonomy of Tor Hidden Services (Spitters et al. 2014).*

### 4.2.1  Markets/Illegal Content

Marketplaces like *Silk Road*, *Hansa* and *AlphaBay* have been popular over the years for allowing trade of questionable goods like illicit drugs, weapons, stolen identities, stolen credit card details, child pornography, and others. Silk Road was one of the first major such marketplaces reaching sales of over USD 1.2 million per month (Christin 2013) mainly consisting of controlled substances and narcotics. Weapons are also a popular commodity on such marketplaces but are overshadowed by the sale of illicit drugs (Rhumorbarbe et al. 2018). Stolen information is also found for sale on these markets as it provides the criminal, and its acquaintances, anonymity. This data can include the sale financial fraud-related documents, e.g. credit card details or identity theft enabling documents like medical records as these typically have a higher shelf life (Hoffman and Rimo 2017).

#### 4.2.2 Communication/Recruitment

The deregulated and anonymous nature of the dark web is of appeal to a range of nefarious actors including terrorist groups and hackers. The dark web allows for anonymous communication activities including recruitment (Weimann 2016a). Terrorist groups can spread their ideology, conduct recruitment, share knowledge, train, advertise, fundraise, target and form communities overall without concern for geographical separation or even the presence of a local leader (Brynielsson et al. 2013; Weimann 2016b). Likewise, the dark web allows for anonymous communication between hackers to share information. Due to the extensive use of dark web forums for such purposes, they have been the target for various forms of monitoring ranging from manual observation to crawling combined with natural language processing techniques for automated threat intelligence and various other insights (Abbasi and Chen 2007).

#### 4.2.3 Cybercrime & Terrorism

Cybercrime or terrorism on the dark web can be enacted by individuals or well-organised groups. Cybercrime is increasingly accessible to anyone who wishes to engage in low-risk criminal activities while still having an impact (e.g. conducting DDoS attacks on websites is as convenient as hiring a botnet which offers DDoS-as-a-Service (Crawley 2016)). These services let malicious actors pick the 'low hanging fruit' (targets without adequate security controls or training). Tor's hidden services can assist in sophisticated attacks, involving command-and-control (C2) channels being maintained between attackers and victims. Tor's offer of anonymity (its difficulty in being shut down) is ideal for C2 servers and this is one of the most popular hidden services available (Biryukov et al. 2014).

Nation states are expressing their concern at their need to prepare for warfare in the digital realm, particularly where this warfare affects Critical Infrastructure (CI) and Industrial Control Systems (ICS) with the potential to have adverse real-world outcomes (Stoddart 2016). This ease-of-entry into being a cybercriminal is further enabled given the asymmetry of the warfare environment. Despite cybercriminals not being as well-funded or resourced as the organisations they target, they possess an advantage on the digital battlefield due to being able to choose their tactics, timing and location whereas the defenders are required to be on alert at all times (Ahmad 2010). In warfare, deterrence and dissuasion have worked as useful military strategies in the past for many reasons, one of them being the high barrier of entry into nuclear weapons. However, this does not apply in cyberspace as a weapon can be coded by people on, or bought easily from, the dark web (Nye 2017). By the same token, deterrence is no longer the domain of governments as sometimes non-state actors can also become active participants in cyber warfare against common enemies (Nye 2017).

Law enforcement agencies and hacktivist groups around the world have been active in the reduction of terrorist groups' activities on the surface web (Weimann 2016a). However, these terrorist groups have migrated to the dark web (Denic 2017; Weimann 2016b) where: a) their supporters can freely express their opinions anonymously, b) their operations can continue to be funded via virtual currencies and c) the dark web serves as a potential recruitment centre and training ground. The latter has been attributed to a significant amount of terrorist activity and identification of this on dark web forums via natural language processing is an ongoing effort (Brynielsson et al. 2013).

#### 4.2.4 Cyber Threat Intelligence

Law enforcement and security product and service providers are regular observers of the cybercrime community on the dark web. Antivirus and other security companies protect their users from malware based on signatures derived from past attacks (Samtani et al. 2017). However, there is a shift towards a more proactive approach to security where Situation Awareness (SA) is integrated into the Information Security Risk Management (ISRM) systems of organisations (Webb et al. 2014), and part of SA is the collection and processing of data which can help with managing security. This idea has given rise to practices in Cyber Threat Intelligence (CTI) which involves gathering data about exploits and cybercriminal activity, particularly in dark web forums (Samtani et al. 2017).

#### 4.2.5 Financial transactions

While the majority of real-world financial transactions are traceable to individuals or entities, the emergence of cryptocurrencies like Bitcoin (Nakamoto 2008) allows for near-anonymous money exchange. This decentralised and distributed form of currency complements the dark web's nature to supply funding for operations without attribution of the source (Broadhurst et al. 2017). Bitcoin offers pseudonymity; all transactions associated with an address are public and can be monitored. However, multiple 'tumbling' services operate on the dark web to effectively launder cryptocurrency, making it difficult to link deposits and withdrawals from individual digital wallets and to link them to specific transactions which may be under investigation (DiPiero 2017). Cryptocurrencies are used to facilitate activity on the dark web for marketplace payments and to fund crime. Ransomware victims are typically presented with the option of paying the attacker in Bitcoins in exchange for having their files decrypted (Sabillon et al. 2016). It is proposed that a well-resourced attacker may be able to compromise the identity of Bitcoin-over-Tor users (Biryukov and Pustogarov 2015), however, with an increasingly distributed nature of financial transactions and the markets themselves, DiPiero (2017) argues that efforts are best utilised in identifying offenders instead of disrupting the technology itself.

### 4.2.6 Proxy to Surface Web

Many civilians also use the Tor darknet infrastructure to browse the surface web, or hidden service versions of popular websites on the surface web e.g. Facebook (social media), ProPublica (news), DuckDuckGo (search engine), and other services which may be restricted locally or provide cause for concern on privacy/ad tracking (Lexie 2018). Residents in China employ it to bypass the 'great firewall of China' internet filter which prevents them from accessing many popular services. Journalists also use the dark web to write about politically sensitive content which may be punishable by the law of their country. Similarly, whistleblowers may communicate with journalists and leak sensitive data to have it published on the surface web, but fear for their safety (Chertoff and Simon 2015).

### 4.2.7 Summary of Roles

Table 1 shows a summary of the roles of the dark web.

| Dark Web Role | Description (with representative citations) |
|---|---|
| As a Market | Trading illicit drugs (Maddox et al. 2016; Rhumorbarbe et al. 2018; Tzanetakis 2018); trading malware and exploits (Ablon et al. 2014); trading credit card, identities & stolen information (Broadhurst et al. 2017; Denic 2017); trading child abuse media (Dalins et al. 2018); trading weapons (Nunes et al. 2016; Rhumorbarbe et al. 2018). |
| As a Communication platform | *Forums* for discussion (Abbasi and Chen 2007; Broadhurst 2017; Sapienza et al. 2017); *Chat* for real-time communication (Maddox et al. 2016; Sabillon et al. 2016; Weimann 2016a). |
| As an enabler of Cybercrime | *Malware-as-a-Service* for criminal services (Tsakalidis and Vergidis 2017); *Command-and-Control (C2)* servers deployed as hidden services (Owen and Savage 2016); *Terrorism Operations* conducted in conjunction with other roles (Denic 2017; Weimann 2016a). |
| As a source of Threat Intelligence | *Scanning Forums & Marketplaces* for threat intel (Nunes et al. 2016; Samtani et al. 2017; Skopik 2017) |
| As an enabler of anonymous Financial Transactions | Using *Bitcoin over Tor* for anonymity (Biryukov and Pustogarov 2015; DiPiero 2017); *Money Laundering* of cryptocurrencies via tumbling services (Dalins et al. 2018; Denic 2017; Moore and Rid 2016; Sabillon et al. 2016). |
| As a Proxy to the Surface Web | *Avoid censorship* by circumventing blocks (Chertoff and Simon 2015; Denic 2017); *Protection from persecution* by local authorities due to browsing anonymity (Chertoff and Simon 2015; Denic 2017; Owen and Savage 2016). |

*Table 1: Roles played by the Dark Web as per literature*

## 4.3 Legal and Societal Concerns

Considerable research is being conducted into finding vulnerabilities in Tor. There is value for law enforcement in being able to deanonymize users of Tor, or at least disrupt communications making the system unusable (Vogt 2017). Whilst law enforcement agencies have been targeting and successfully shutting down dark web marketplaces (such as *Silk Road*, *AlphaBay* and *Hansa* (Tzanetakis 2018)), it does not take long for the marketplaces to return to a functional form. (Maddox et al. (2016) report that a new version of Silk Road continues to exist despite takedowns). Furthermore, fully distributed marketplaces, ones without a need for servers to host the platform, are becoming mature in recent years (e.g. OpenBazaar). Combined with a distributed currency system like Bitcoin and the Tor protocol such market systems can afford near anonymity to its users and their engagements.

To achieve deanonymisation, there are passive methods that work by passively fingerprinting the circuits set up in Tor connections (Kwon et al. 2015), analysing traffic within middle relays (Jansen et al. 2017) or by correlating traffic between the entry and exit nodes (Overlier and Syverson 2006). If adversaries operate at the nation-state level and are able to influence entire autonomous systems (Sun et al. 2015) they may be able to deanonymize nearly 100% of Tor traffic within months of engagement (Johnson et al. 2013). Users may also be deanonymized by having their identities linked across multiple hidden services by analysing their writing style (*stylometry*) (Ho and Ng 2016). Denial of Service is also explored as a method of discouraging use or supplementing deanonymization efforts above by forcing users to connect to relays under the attacker's control (Jansen et al. 2014).

Despite the persistence of law enforcement, developers and maintainers of Tor continually upgrade the protocol and software to fix vulnerabilities and introduce new functionality to make the system even safer to use (Winter 2017). Similar networks are also becoming more widely adopted (e.g. the P2P technology behind Freenet is known to be more secure than Tor (Duddu and Samanta 2018)).

## 5 Research Agenda

In this section, we discuss the areas in which we have identified gaps in the current Dark Web literature. We conclude each of the areas with a set of one or more future research questions to set up the research agenda. These research questions are aimed at filling the gaps in current Dark Web research.

### 5.1 Enabling Cybercrime: Criminals' Dependence on the Dark Web

Cybercriminals have benefitted from hidden services by buying exploits (including zero-day) from marketplaces, engaging with other criminals on forums, hosting command-and-control (C2) servers for botnets & exfiltration malware, and hiring professional criminals or their skills 'as-a-service' (Chertoff and Simon 2015; Sabillon et al. 2016; Vogt 2017). The anonymity attribute that the dark web confers to its users makes these activities possible and enables cybercriminals to conduct their affairs.

Various types of cybercrime and actors are present, ranging from amateur 'script-kiddies' using readily-available tools and exploits through to highly organised Advanced Persistent Threat (APT) actors which are known to have caused significant cases of espionage against large organisations and nation states (Ahmad et al. 2019). These attacks target digital assets in organizations (data, algorithms, models, analytics, systems, and infrastructure) (Shedden et al. 2016; Shedden et al. 2011). No significant literature was found explicitly exploring a link between the dark web and APT phenomena, and if investigated it can potentially be exploited in curbing or better defending against cyber-attacks from APTs in the future.

*Future RQ1.    To what extent is the dark web crucial to the success of APT attacks?*

### 5.2 Distributed Hidden Services & Dark Web Scalability

Hidden services on the dark web make use of anonymity to keep their location and address private. This makes it challenging for law enforcement to shut down these services as their hosting provider and geographical location are hidden. However, as seen recently, darknet marketplaces like Silk Road, Hansa and AlphaBay are not impervious to regulation (Broadhurst 2017; Tzanetakis 2018). As a result, marketplace operators are exploring ways of protecting themselves further. In these situations when anonymity is not enough, a more distributed implementation is potentially the answer and can be seen in its early stages in the form of multiple new decentralised markets either built on Tor (Fraser 2015) or blockchain technology (Silkos (SLK) 2017).

Research can be conducted into the scalability of the dark web to determine whether current internet infrastructure continues to support this phenomenon or is a limiting factor in its usage. Moreover, given that some methods for deanonymization of clients or services rely on controlling sufficient nodes on the Tor network (Jansen et al. 2014; Johnson et al. 2013; Kwon et al. 2015), further investigation can be conducted into what level of growth, or user adoption, of Tor is required before these methods become economically infeasible to exploit. This research could help law enforcement reprioritise its approaches for monitoring the dark web.

*Future RQ2.    What trends can be observed about existing web infrastructure, especially hidden services, moving in the direction of being more decentralised?*

*Future RQ3.    What technological, and other, factors currently affect this transition and how will this affect the scalability and observability of the dark web?*

### 5.3 Role & Capability of Law Enforcement

As technology advances, it is becoming increasingly difficult to track, identify and shut down sites allowing illegal activity, especially as they cease to exist in their current form (Moore and Rid 2016). The dark web is allowing communication between people to become more decentralised and thus harder to regulate, which may allow for a proliferation of unethical (and illegal) content whilst increasing the difficulty of monitoring by law enforcement. Subsequently, a major focus of law enforcement is on the deanonymization of hidden services, of marketplace operators, of site administrators and anyone else involved in the facilitation of the relatively centralised (but anonymous) infrastructure used to conduct illegal activity on the dark web.

*Future RQ4.*   What strategies can law enforcement use to exercise control over the dark web?

*Future RQ5.*   How can society reduce unethical or criminal behaviour in the presence of a market demand that seems to be forever present combined with markets that are difficult to shut down?

Furthermore, the questions of what *can* be done versus what *should* be done about the technology from a regulatory perspective are mutually exclusive in their pure forms. Promoting the full extent of civil liberties by way of allowing unrestricted and unmonitored access to internet technologies mean illicit activities will likely flourish (Chertoff 2017). However, civil liberties may suffer when policies restricting the use of the surface, or dark, web are brought into existence. It is argued the focus should be on initiatives elsewhere in the process (e.g. prosecuting the traders as opposed to facilitators of the marketplace) as this becomes more and more difficult to do with technological advancements. DiPiero (2017) suggests increasing the number of sting operations and focusing on bringing down the source of drugs rather than the marketplace itself as the phenomenon grows over time.

*Future RQ6.*   How can legal jurisprudence strike the right balance between safeguarding civil liberties and sanctioning unethical behaviour?

## 5.4   Dark Web Phenomenon Growth vs Regulatory Growth

Most research on the topic of the dark web focuses on deanonymization or utilising it for some form of threat intelligence. This topic, as a manifestation of the phenomenon of civilians looking for private spaces to conduct their affairs, is of research interest. Seeing the dark web through this lens could allow for better speculation around its growth and may help stakeholders (such as law enforcement) make better investment decisions. We argue that the dark web has increased in size as regulations have increased and individual liberties have decreased (Shor et al. 2018). The use of the dark web by cyber criminals to facilitate their operations and the resulting efforts by law enforcement on curtailment could lead to investment and advancements in the dark web phenomenon (for example Mexican drug cartels are highly advanced and technologically innovative which drives growth (Ramirez and Bunker 2014)). Clamping down on illegal activity has become the driver for technological innovation and growth in the direction suited for illegal activity, especially when the economic return from said activities is able to support this investment.

Potential for research exists into whether the phenomenon of the dark web (and investment in technology which enables it) is increasing over time and if this increase is correlated with a reduction in personal liberties or stricter regulations. Establishing this correlation leads to a discussion on the effects of policy concerning civil liberties on the dark web phenomenon, and a desired outcome in the latter could be affected by manipulating the former.

*Future RQ7.*   What is the role of regulation in moderating the growth of the dark web and investment in its technology in the context of personal, civic and economic liberties in a society?

## 6   Conclusions

In this paper, we conduct a systematic literature review into the roles the dark web plays in modern digital society, its enablement of cybercrime and its relationship with law enforcement and society. Conducting a literature review on this topic was challenging as information available and research conducted on the topic is relatively limited due to its 'secretive' and non-identifying nature.

The first research question required investigation of the roles the dark web currently serves. While online marketplaces for illicit drugs are still a major use, it remains one of many that are available ranging from marketplaces for trade of software exploits through to extremist recruitment. The second research question involved looking at the significance of the dark web to cybercriminal activities and operations. Evidence shows that the dark web serves as a tool for many criminal activities such as training ground for cybercriminals (e.g. APT actors) via discussion boards and hosting infrastructure used to conduct, support and commodify cybercriminal operations. However, in real terms, the dark web only lowers the barriers of entry into the world of cybercrime, as more established and well-resourced actors are purported to use their own private infrastructure (e.g. botnet proxies) to anonymise their actions, discovering their own zero-day vulnerabilities and using other means to finance their operations. Finally, the third research question was about law enforcement curbing illegal activity on the dark web. Law enforcement continues in its attempts to close markets responsible for the sale of illicit items, especially recreational drugs. However, the infrastructure reacts to these incidents and attempts to mitigate the vulnerabilities exploited by the authorities. In response, the Tor protocol is continually updated to fix protocol vulnerabilities, market operators improve their operational security, the financial backbone (cryptocurrencies) introduces newer and more anonymous variants alongside an increase in money laundering ('tumbling') services, hidden services become more distributed, and escrow functions are evolving to be based not just on trust but also on cryptographic improvements.

In the debate on privacy versus security, the technology factor is weighing in stronger than before as it becomes not just a matter of legality but of technical capability to monitor and conduct surveillance on people. It is possible that in the future this debate would be over as technology advances to the point where privacy is not at the mercy of governments but in the hands of users and private corporations. The discussion on this topic then no longer remains in the domain of technology but is one that needs an interdisciplinary contemplation by experts in areas of psychology, sociology, law, and others.